\documentclass[aps,pr,twocolumn,showpacs,superscriptaddress,groupedaddress,nofootinbib]{revtex4}  

\usepackage{longtable}
\usepackage{graphicx}  
\usepackage{ulem}   
\usepackage{comment} 
\usepackage{datetime}
\usepackage{dcolumn}

\usepackage{amsxtra}
\usepackage{euscript} 
\usepackage{bm}        
\usepackage{tabularx}
\usepackage{float}
\restylefloat{table}

\usepackage[cmtip,arrow]{xy}
\usepackage{pb-diagram,pb-xy}

\usepackage{amsmath}
\usepackage{amssymb}
\usepackage{amsthm}
\usepackage[pdftex]{color}
\usepackage[pdftex,colorlinks,citecolor=blue,linkcolor=blue,urlcolor=blue]{hyperref} 
\usepackage{graphicx}
\usepackage{dcolumn} 
\usepackage{bm} 

\usepackage{longtable}

\usepackage{ulem}   
\usepackage{comment} 
\usepackage{datetime}

\usepackage{tabularx}
\usepackage{float}
\restylefloat{table}

\newcommand{\beq}{\begin{equation}}
\newcommand{\eeq}{\end{equation}}
\newcommand{\bea}{\begin{eqnarray}}
\newcommand{\eea}{\end{eqnarray}}
\newcommand{\barr}{\begin{array}}
\newcommand{\earr}{\end{array}}

\long\def\begincomment#1\endcomment{}

\newcommand{\tr}{\mathop{\mathrm{tr}}}

\newtheorem{definition}{Definition}

\newtheorem{proposition}{Proposition}

\usepackage{pgf,tikz}
\usetikzlibrary{arrows}

\usepackage{bm}
\usepackage{bbold}

\usepackage{mathtools}

\DeclarePairedDelimiterX\braket[2]{\langle}{\rangle}{#1 \delimsize\vert #2}

\newcommand{\D}{\mathfrak{d}}

\begin{document}


\title{   Noether's currents for conformable fractional scalar field theories}

\author{Jean-Paul Anagonou}
\email{jeanpaul.anagonou@uac.bj}
\affiliation{International Chair in Mathematical Physics and Applications (ICMPA-UNESCO Chair), University of Abomey-Calavi,
072B.P.50, Cotonou, Republic of Benin}

\author{Vincent Lahoche} \email{vincent.lahoche@cea.fr}
\affiliation{Université Paris Saclay, CEA List, Gif-sur-Yvette, F-91191, France}

\author{Dine Ousmane Samary}
\email{dine.ousmanesamary@cipma.uac.bj}
\affiliation{Université Paris Saclay, CEA List, Gif-sur-Yvette, F-91191, France}

\affiliation{International Chair in Mathematical Physics and Applications (ICMPA-UNESCO Chair), University of Abomey-Calavi,
072B.P.50, Cotonou, Republic of Benin}

\date{\today}

\begin{abstract}
The construction of fractional derivatives with the right properties for use in field theory is reputed to be a difficult task, essentially because of the absence of a unique definition and uniform properties. The conformable fractional derivative introduced in 2014 by  Khalil et al. in their seminal paper is a novel and well-behaved definition of fractional derivative for a function which is derivable in the usual sense. 
In this paper, we investigate the consistency of the Euler-Lagrange formalism for a field theory defined on such a fractional space-time. We especially focus on the relation between symmetries and conservation laws (Noether's currents), about the symmetry group introduced to construct the Lagrangian of the field. In particular, we show that the use of the conformable derivative induces additional terms in the calculation of the action variation. We also investigate the conservation of the Noether current and show that this property only takes place on condition that the equations of motion are verified with a new definition of the conserved law. 

\end{abstract}


\maketitle


\section{Introduction}
Since the pioneering work by Emmy Noether \cite{Noether}, which established a strong connection between symmetries and conservation laws, the investigation of Noether currents has become a classical topic in the study of the properties of classical and quantum physical systems (see \cite{Rosen:1972ku}-\cite{LucioMartinez:1997pa} and references therein), especially for discussing integrability. 
 The generalization of Noether's theorem for physical theories beyond standard paradigms is then an important topic, and the aim of this paper is essentially to consider it for a field theory with fractional derivatives, defined on the space of $\alpha$-differential functions. There are many motivations for this type of space. Already at the most elementary level, in the formalism of path integrals, the “smooth” character of classical trajectories emerges from purely Markov processes, not differentiable in the ordinary sense. Furthermore, this is certainly part of the parallel but complementary works aimed at finding a good definition and formulation of quantum gravity. This generalization may also help to deduce new properties that we might otherwise miss in classical treatment \cite{Baez}. 
\medskip

The appearance of the fractional derivative in this context makes field theory non-local, also due to the deformation of the d'Alembert operator of the kinetic part of the functional action \cite{Calcagni:2022shb}. Note that in general, where spacetime is deformed such as noncommutative spacetime, the corresponding Noether current, coming from the symmetry of the functional action is not locally conserved \cite{Gerhold:2000ik}. This pathology is repeated in the case of fractional derivatives and will be discussed largely in this paper.

Let us briefly recall historically the origin of the fractional derivative. This origin can be traced to the 17th century, with the exchanges between Leibniz and de l'Hospital. But it was Joseph Liouville who in 1832 laid the first foundations of the fractional analysis.
The history of this branch of mathematics has been comparatively less prolific than that of integer derivatives.
The reason for this, despite the lack of experimental proof of the fractional nature of space at this date, is the absence of an unequivocal definition of the fractional derivative. Even worse, these definitions can be in contraction one with the other, even if they are all equally legitimate \textit{a priori}. 
We thus speak of derivatives in the sense of Louiville-Riemann, Liouville-Weyl, Caputo or Grünwald-Letnikov \cite{Tarasav}. Moreover, when they exist, it is customary today to define these fractional derivatives from the Fourier or Laplace transforms of the functions. It was not until the work of Laurent Schwartz on the theory of distributions in the middle of the 20th century that most of these definitions were agreed upon. Indeed, and surprisingly, it is through the theory of integration that fractional derivatives find their legitimacy and internal coherence. Indeed, integration is a more flexible concept, which can also be defined for non-derivative functions such as the Weierstrass function, whose curve is of fractal dimension and derivable nowhere, but which is nevertheless integrable. Thus, the integral was and still is used as a support for the current definitions of the fractional derivative \cite{Khalid}-\cite{Podlubny}.

The study of fractional derivatives has recently taken a new momentum in the scientific community, and in particular in physics. It has numerous applications in solving problems in spaces where classical laws are no longer in agreement with empirical results, such as non-differentiable spaces, fractal objects etc \cite{Nottale:2012cb}-\cite{Nottale:2009azf}. study of fractional derivatives has recently taken momentum in the scientific community and in particular in physics because of its numerous applications in solving problems in spaces where classical laws are  These exotic objects find themselves attached to the most frontier topics of our current knowledge, such as quantum gravity where the usual concepts of space, time, and continuity disappear, and where the usual mathematics of differential varieties becomes inoperative. The study of spacetime defects is also an accepted framework for the use of such mathematical objects \cite{Carqueville:2023jhb}-\cite{Vachaspati:1991tr}.

Especially because of its intrinsically non-local character, the fractional derivative finds applications in domains characterized by strong non-localities as is the case for example in high energy physics, in many modern approaches to quantum gravity. Links have for instance been established between fractional geometry and non-commutative space-time, and fractional differential equations are a natural framework to define a scattering process behaving in $x(t)\sim t^{1/\alpha}$ when $\alpha \in [0,1[$, see \cite{Tarasav} and references therein. Finally, as soon as the modelling of physical phenomena involves a “long” memory, moving away from a Markovian process, requiring the introduction of integrodifferential terms, modelling by fractional dynamics appears natural. This is the case for example in material physics for linear viscoelasticity problems with a long memory, viscous-thermal phenomena in acoustics, or in polymer physics \cite{Batarfi}-\cite{Schneider}. This offers only a very narrow perspective on this vast and exploding subject. 

In this paper, we are aiming to investigate a field theory on a fractional background spacetime, which requires the choice of a definition of the derivative generally not obvious as space dimension is not necessarily integer. We construct the EL approach for the classical dynamics of scalar fields defined on a fractional space, and to investigate conservation laws. Because a definition of derivative is required, we consider the Khalil $\&$ \textit{al.} definition given in ref \cite{Khalid} and \cite{Thabet} called \textit{conformable fractional derivative}. More precisely, we consider \textit{left} and \textit{right} derivatives, considered in \cite{Thabet} and defined as follows:
\begin{definition}\label{def1}
Let $f$ be a differential real function in the usual sense and $\alpha \in ]0,1]$. For $t\in [a , + \infty [,\, a\in\mathbb{R}$, the order $\alpha$ conformable fractional derivative is defined as: 
\bea
_{}^a\D^{\alpha}_t f(t) = \lim_{\epsilon \longrightarrow 0}\frac{f(t+\epsilon(t-a)^{1-\alpha}) - f(t)}{\epsilon}. 
\eea
In the same manner, for $t\in [-\infty , b ], \, b\in\mathbb{R}$: 
\bea
\D^{\alpha, b}_t f(t) = -\lim_{\epsilon \longrightarrow 0}\frac{f(t+\epsilon(b -t)^{1-\alpha}) - f(t)}{\epsilon}.
\eea
We also assume that the conformable derivative at the points $a$ and $b$ exist, and we denote it by $_{}^a\D^{\alpha}_t f(a):=f_R^{(\alpha)}(a) $ and $\D^{\alpha, b}_t f(b):=f_L^{(\alpha)}(b) $ respectively.
\end{definition}
Note that our definition also proposed in reference \cite{Thabet} is motivated by the definition in \cite{Khalid} in the hope of extending it to functions with negative arguments. Also, the parameters $a$ and $b$ will be set to zero to cover the whole space and have no additional implication in our definition. The conformable derivatives in the definition \eqref{def1} are
 called $\alpha$-derivative.
Unlike the fractional derivative of Riemann-Liouville and Caputo, this derivative has better properties such as Leibniz's rule, thus allowing good use in field theory and therefore has technical advantage defined by a limit, which is more appropriate for a variational approach to dynamics. To be more precise about the physical difficulty which appears with another fractional derivative, see \cite{Rami}-\cite{Tarasov} and reference therein.

The paper is organized as follows: In section \eqref{sec2}, we construct the theoretical ingredients that allow us to construct the Lagrangian formalism in the context of conformable fractional. The EL equation of motion is then given explicitly. derivative. In section \eqref{sec3} the generalization of Noether theorem is derived, and we also give a consistent proof of our results to be explicit.  Section \eqref{sec4} is devoted to an application of our study to a very simple system (the harmonic oscillator in one dimension). The non-local conservation of the energy of the system which is the first component of the energy-momentum tensor (EMT) is also studied as well as its regularization. In section \eqref{sec5} we provide the conclusion of our work and announce our forthcoming investigation.

\section{Field theory with conformable derivative}\label{sec2}
This section aims to propose a Lagrangian formalism for the dynamics of a (Euclidean) field described in terms of fractional derivatives. The classical field is assumed to be a smooth function $\phi:\mathbb{R}^D\longrightarrow \mathbb{R}$,
\begin{equation}
\bm{x}=(x^1,\cdots,x^D) \to \phi(\bm x)\,,
\end{equation}
and we call arbitrarily ‘‘time" the first coordinate $x_1$. A field theory requires a definition of the partial derivative:

\begin{widetext}
  
\begin{definition}\label{def2}
We consider a smooth function $\phi$ defined on  the Euclidean vector space  $\mathbb{R}^D$ as $\phi:\mathbb{R}^D\longrightarrow \mathbb{R}$. For $ { \bm x}= (x^i, {\bm x}_{\bot_i}) := (x^1,\cdots, x^i,\cdots,x^D)\in\mathbb{R}^D$ with ${\bm x}_{\bot_i} = (x^1, \cdots, x^{i-1}, x^{i+1},\cdots, x^D)\in\mathbb{R}^{D-1}$ and   ${\bm a}= (a^i, {\bm a}_{\bot_i})\in\mathbb{R}^D$,
 the right fractional partial derivative with respect to the $i^{th}$ coordinates is given  for $x^i\in [a^i,\infty[$ by:
\bea\label{}
_{}^a\D_i^{\alpha }\phi(x^i, {\bm x}_{\bot_i}) &=& \lim_{\epsilon \longrightarrow 0}\frac{\phi(x^i + \epsilon (x^i - a^i)^{1-\alpha}, {\bm x}_{\bot_i}) - \phi(x^i, {\bm x}_{\bot_i})}{\epsilon}.
\eea
Also For $ { \bm x}= (x^i, {\bm x}_{\bot_i}) \in\mathbb{R}^D$ and ${\bm b}= (b^i, {\bm b}_{\bot_i})\in\mathbb{R}^D$, the left fractional partial derivative with respect to the $i^{th}$ coordinates is given by for $x^i\in]-\infty,b^i]$: 
\bea\label{}
\D_i^{\alpha, b}\phi(x^i, {\bm x}_{\bot_i}) &=& -\lim_{\epsilon \longrightarrow 0}\frac{\phi(x^i + \epsilon (  b^i - x^i)^{1-\alpha}, {\bm x}_{\bot_i}) - \phi(x^i, {\bm x}_{\bot_i})}{\epsilon}.
\eea
where $\alpha\in ]0,1]$. 
\end{definition}
\end{widetext}
This definition generalizes slightly the  Khalid's $\&$ \textit{al.} definition given in definition \ref{def1}. Note that, because $f$ is assumed to be derivable in the definition \ref{def1}, the conformable derivative is related to the ordinary derivative as:
\bea\label{i1}
&&_{}^a\D_i^{\alpha }\phi({\bm x}) = (x^i-a^i)^{1-\alpha} \partial_i \phi({\bm x}).\\
&&\D_i^{\alpha, b}\phi({\bm x})=-(b^i-x^i)^{1-\alpha}\partial_i \phi({\bm x}).\label{corresp}
\eea
where $\partial_i:=\frac{\partial}{\partial x^i}$ is the ordinary derivative with respect to the coordinates $x^i$. 

If there is no ambiguity, we will use the notation $\D_i^\alpha$ to indicate the left or right derivative having identical properties at given times.
It has to be noticed that conformable derivative satisfies standard axioms of derivative:
\begin{enumerate}
    \item Leibnitz rule:
\begin{equation}
\D_i^\alpha(\phi_1\phi_2)({\bm x})=\phi_1({\bm x})\D^\alpha_i \phi_2({\bm x}) + \phi_2({\bm x})\D^\alpha_i \phi_1({\bm x})\,,
\end{equation}
\item Linearity ($\forall\,\mu,\nu \in \mathbb{R}$):
\begin{equation}
\D_i^\alpha(\mu\phi_1+\nu\phi_2)({\bm x})=\mu \D_i^\alpha\phi_1+\nu \D_i^\alpha \phi_2\,,
\end{equation}
\item Chain rule:
\begin{equation}
\D_i^\alpha (f(g(\bm x))= f^\prime(g)\vert_{g=g(x)} \D_i^\alpha g(\bm x)\,.
\end{equation}
\end{enumerate}
We are aiming to construct explicitly EL equations from a variational principle.
Note that the construction of such a variational principle agrees with the definition \ref{def1} of the conformable derivative as a limit. Our starting point is therefore the assumption that there is a functional $\mathcal{S}^\alpha[\phi]=\mathcal{S}_R^{\alpha }[\phi]+\mathcal{S}_L^{\alpha }[\phi]$ of the trajectory that we call action, such that physically relevant trajectories make the action stationary. $\mathcal{S}_L^{\alpha }[\phi]$ is a part of the action constructed with the left fractional derivative, i.e. in the negative spatial domain and $\mathcal{S}_R^{\alpha }[\phi]$ is a part of the action constructed with the right fractional derivative, i.e. in the positive spatial domain. Furthermore, we assume the existence of a function that we call Lagrangian density $\mathcal{L}=\mathcal{L}_L+\mathcal{L}_R$, depending only on the field and its first $\alpha$-(partial) derivatives and such that $\mathcal{L}$ depends on the parameter $\alpha$ only through the $\alpha$-derivatives. Finally, the classical action $S^\alpha[\phi]$ is assumed to be related to the Lagrangian density as:
\begin{equation}\label{mamequan}
 \mathcal{S}_R^\alpha [\phi]=  \int_{\mathcal{D}_R^{\bm a}} \,\Delta_{R}^{\alpha}({\bm x}) \,d^D{\bm x} \,\mathcal{L}_R[\phi, _{}^a\D_1^{\alpha }\phi,\cdots,_{}^a\D_D^{\alpha }\phi]\,.
\end{equation}
\begin{equation}\label{mamequann}
 \mathcal{S}_L^\alpha [\phi]=  \int_{\mathcal{D}_L^{\bm b}} \,\Delta_{L}^{\alpha}({\bm x}) \,d^D{\bm x} \,\mathcal{L}_L[\phi, \D_1^{\alpha,b }\phi,\cdots,\D_D^{\alpha,b }\phi]\,.
\end{equation}
where $\mathcal{D}_R^{\bm a}:=\times_{i=1}^D[a^i,+\infty[$ and $\mathcal{D}_L^{\bm b}=\times_{i=1}^D]-\infty, b^i]$. Remark also that, to recover entirely the domain $\mathbb{R}^D$
 the limit ${\bm a}\rightarrow {\bm b}$ automatically applies. Furthermore, note that we focus on trajectories derivable in the ordinary sense, such that equations \eqref{corresp} make sense. The factors $\Delta_{R}^{\alpha}$ and  $\Delta_{L}^{\alpha}$ are  the first source of $\alpha$-dependency of  $\mathcal{S}^\alpha[\phi]$ that does not come from derivatives, explicitly:
\begin{equation}
\Delta_{R}^{\alpha}({\bm x}):=\prod_{i=1}^D (x^i-a^i)^{\alpha-1},\,\, \Delta_{L}^{\alpha}({\bm x}):=\prod_{i=1}^D (b^i-x^i)^{\alpha-1}\,.
\end{equation}
The origin of this factor comes from the requirement that integration is the inverse operation of a derivative. Concretely, for a function $f(t)$ of a single variable, we define the primitive $I(f)$ using the definition \eqref{def1} as:
\bea
_{}^aI^\alpha(f)(s)=\int_a^s\Delta_{R}^\alpha(t) dt\, f(t)\\ I^{\alpha,b}(f)(s)=\int_{s}^b \Delta_{L}^\alpha(t) dt\, f(t).
\eea
And therefore the above definition is generalized to functions on $\mathbb{R}^D$ in the following forms:  
\bea
_{}^aI^\alpha_i \phi({\bm x}) = \int_{a^i}^{x^i} \frac{ \phi({\bm y})}{(y^i - a^i)^{1-\alpha}}dy^i,
\eea
\bea
I^{\alpha, b}_i \phi({\bm x}) = \int_{x^i}^{b^i} \frac{ \phi({\bm y})}{(b^i - y^i)^{1-\alpha}}dy^i \quad 
\eea
such that:
\bea\label{ye}
_{}^a\D^\alpha_i \circ _{}^aI^\alpha_i \phi({\bm x}) &= &_{}^aI_i^\alpha\circ _{}^a\D_i^\alpha \phi({\bm x})\,\,:=\,\,\phi({\bm x})  
 \eea
\bea\label{yee}
\D^{\alpha, b}_i \circ I^{\alpha, b}_i \phi({\bm x})&=&  I_i^{\alpha, b}\circ \D_i^{\alpha, b} \phi({\bm x})\,\,:=\,\,\phi({\bm x}).
\eea
The full integration taking into account all the coordinates will be 
\bea
_{}^aI^\alpha\phi({\bm x})=\int_{\bm a}^{\bm x}\Delta_R^\alpha({\bm y})\phi({\bm y})d^D{\bm y}
\eea
\bea
I^{\alpha,b}\phi({\bm x})=\int^{\bm b}_{\bm x}\Delta_L^\alpha({\bm y})\phi({\bm y})d^D{\bm y}
\eea

\begin{figure}
\begin{center}
\includegraphics[scale=0.8]{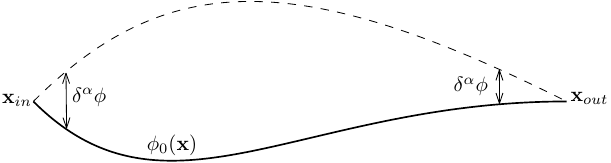}
\end{center}
\caption{Path in the fractional spacetime with fixed boundary points $\delta^\alpha({\bm x}_{in})=\delta^\alpha\phi({\bm x}_{out})=0$}\label{figBubbles}\label{fig1}
\end{figure}
Let's now move on to the construction of a dynamic that is compatible with the previous definitions. Around a typical trajectory $\phi_0(\bm x)$, we construct a small variation $\phi_0\to \phi_0+\delta^\alpha \phi$, such that $\delta^\alpha \phi$ vanishes outside and on the boundary of some $D$-dimensional submanifold $M_D\subset \mathbb{R}^D$, $\partial M_D\neq \emptyset$ (figure \ref{fig1}). At first order, the variation of the action reads: $\delta^\alpha \mathcal{S}^\alpha=0$ which is explicitly written as 
\bea
_{}^a\delta^\alpha \mathcal{S}_R^\alpha= \int_{\mathcal{D}_R^{\bm a}} \,\Delta_{R}^\alpha({\bm x}) d^D{\bm x} \,_{}^a\delta^\alpha\mathcal{L}_R[\phi, _{}^a\D^{\alpha}\phi]=0,\\
\delta^{\alpha,b} \mathcal{S}_L^\alpha= \int_{\mathcal{D}_L^{\bm b}} \,\Delta_{L}^\alpha({\bm x}) d^D{\bm x} \,\delta^{\alpha,b}\mathcal{L}_L[\phi, \D^{\alpha,b}\phi]=0
\eea
with the initial condition $\phi({\bm a})=0=\phi({\bm b})$ and where we set $\mathcal{L}_R[\phi, _{}^a\D_1^{\alpha}\phi,\cdots, _{}^a\D_D^{\alpha}\phi]:=\mathcal{L}_R[\phi,  _{}^a\D^{\alpha}\phi]$ and $\mathcal{L}_L[\phi, \D_1^{\alpha, b}\phi,\cdots, \D_D^{\alpha, b}\phi]:= \mathcal{L}_L[\phi,  \D^{\alpha, b}\phi]. $
The variation of the Lagrangian density is:
\begin{align}
& _{}^a\delta^\alpha \mathcal{L}_R[\phi, _{}^a\D^{\alpha}\phi ] = \partial_\phi\mathcal{L}_R[\phi, \,_{}^a\D^{\alpha}\phi\,]\, _{}^a\delta^\alpha \phi \cr
&+ \sum_{i = 1}^{D}\Big( \partial_{_{}^a\D_i^{\alpha}\phi} \mathcal{L}_R[\phi, \,_{}^a\D^{\alpha}\phi ] \,_{}^a\delta^\alpha (_{}^a\D_i^\alpha \phi) \Big),
\end{align}
 and 
\begin{align}
 &\delta^{\alpha, b} \mathcal{L}_L[\phi, \D^{\alpha, b}\phi ] = \partial_\phi\mathcal{L}_L[\phi, \D^{\alpha, b}\phi\,]\delta^{\alpha, b} \phi \cr
 &+ \sum_{i = 1}^{D}\Big( \partial_{\D_i^{\alpha, b}\phi} \mathcal{L}_L[\phi, \D^{\alpha}\phi ] \delta^{\alpha, b} (\D_i^{\alpha, b}\phi) \Big).
\end{align}
By definition the operator $\delta^\alpha$ commute with $\alpha$-derivative in the left and right case, $\delta^\alpha(\D^\alpha\phi)=\D^\alpha(\delta^{\alpha}\phi)$, we have:
\begin{align}
_{}^a\delta^\alpha \mathcal{S}_R^\alpha[\phi] 
&=\int_{\mathcal{D}_R^{\bm a}}\,\Delta_{R}^\alpha ({\bm x})d^D{\bm x}\Bigg\{ \partial_\phi\mathcal{L}_R\,  \cr
&-  \sum_{i = 1}^{D}\, _{}^a\D_i^\alpha \Big( \partial_{ _{}^a\D_i^{\alpha}\phi}\mathcal{L}_R \Big)\,\Bigg\}\,(_{}^a\delta^\alpha \phi)+\mathcal{Q}_R^\alpha,
\end{align}
where the extra term denoted by 
$\mathcal{Q}_R^\alpha$,
\begin{equation}
\mathcal{Q}_R^\alpha := \sum_{i = 1}^{D}\int_{\mathcal{D}_R^{\bm a}}  \Delta_{R}^\alpha ({\bm x})d^D{\bm x} \Big[ \,_{}^a\D_i^\alpha\Big( \partial_{\, _{}^a\D_i^{\alpha}\phi} \mathcal{L}_R\,  _{}^a\delta^\alpha \phi\Big) \Big]\,,
\end{equation}
looks as an exact divergence because of the definition of $\Delta_{R}^\alpha({\bm x})$, and vanishes thanks to the standard Green-Ostrogradski theorem. This term contributes to the current (see the next section) and vanishes 
due to the boundary condition. Finally, the EL equation reads
\begin{equation}\label{xxxx}
\partial_\phi\mathcal{L}_R - \sum_{i = 1}^{D}  (x^i - a^i)^{1-\alpha}\partial_i \big( \partial_{\, _{}^a\D_i^{\alpha}\phi}  \mathcal{L}_R \big) = 0\,.
\end{equation}
Using the same analysis we also get for the left fractional derivative, the EL equation
\beq\label{xxxx1}
\partial_\phi\mathcal{L}_L + \sum_{i = 1}^{D}  ( b^i - x^i)^{1-\alpha}\partial_i \big( \partial_{ \D_i^{\alpha, b}\phi}  \mathcal{L}_L \big). = 0
\eeq
We will come back to these equations (\eqref{xxxx} and \eqref{xxxx1}) and try to understand them better by limiting ourselves to the case of a one-dimensional harmonic oscillator.

\section{Generalized Noether's theorem}\label{sec3}

Let $\mathcal{G}$ some Lie group of dimension $M$. We denote as  $\mathcal{T}_\mathcal{G}^x$ and $\mathcal{T}_\mathcal{G}^\phi$ the representations respectively on coordinates and fields. More concretely, we assume that there exist some functions $\{f^i\}$ and $F$, depending on a family of parameters $\vec{\beta}=(\beta_1,\cdots,\beta_M)\in \mathbb{R}^M$, such that:
\begin{equation}\label{tj}
\begin{cases}
\mathcal{T}_\mathcal{G}^x: x^i &\longrightarrow x^{\prime i} =  f^i(\bm x, \vec\beta)\\
\mathcal{T}_\mathcal{G}^\phi:\phi({\bm x})&\longrightarrow \phi'({\bm x}) = F(\phi({\bm x}),\bm x, \vec\beta)\,,
\end{cases}
\end{equation} 
the parameters $\vec{\beta}$ being such that for $\vec{\beta}=0$ the transformation reduces to the identity: $f^i(\bm x, \vec0)=x^i$ and $F(\phi({\bm x}),\bm x, \vec 0)=\phi(\bm x)$. Because $\mathcal{G}$ is a Lie group, we can construct transformations in the vicinity of the identity. We denote them as $\mathcal{T}_I^x$ and $\mathcal{T}_I^\phi$, 
\beq\label{t}
\begin{cases}
\mathcal{T}_I^x: x^{i} \longrightarrow x^{\prime i} = x^{i} - \sum_{\sigma=1}^{M} f^{i}_{(\sigma)}({\bm x})\,\beta^{(\sigma)} \\
\mathcal{T}_I^\phi: \phi({\bm x})\longrightarrow \phi'({\bm x}) = \phi({\bm x}) + \sum_{\sigma=1}^{M} C_{(\sigma)}({\bm x},\phi(\bm x))\,\beta^{(\sigma)}\,.
\end{cases}
\eeq
We say that $\mathcal{G}$ is a symmetry group if it leaves the action unchanged, and we come to the following statement:
\begin{proposition}\label{prop1}(Noether theorem) Assuming that the action \eqref{mamequan} is invariant under $\mathcal{T}_\mathcal{G}$ i.e. $\mathcal{T}_\mathcal{G} \mathcal{S}_R^\alpha-\mathcal{S}_R^\alpha=0$ for all $x^i\geq a^i$. Then there exists a globally conserved vector current given by
\bea\label{CourantR}
\Theta^{\alpha, i}_{(\sigma)R} &=&  C_{(\sigma)}({\bm x}) \partial_{\,_{}^a\D_i^{\alpha}\phi}\,\mathcal{L}_R +  \sum_{k =1}^{D}f^k_{(\sigma)}  \partial_k \phi({\bm x})\,  \partial_{\,_{}^a\D_i^{\alpha}\phi}\,\mathcal{L}_R \cr
&&-  (x^i - a^i)^{\alpha -1} f^{i}_{(\sigma)}({\bm x})\,_{}^a\mathcal{L}_R
\eea
In the case where the transformation $\mathcal{T}_I^x$ is the infinitesimal translation of vector ${\bm \epsilon}$ this current is reduced to the EMT
\bea\label{EMT}
\mathcal{T}^{\alpha, i}_{ j, R} = -  \partial_j \phi({\bm x})\, \partial_{\,_{}^a\D_i^{\alpha}\phi}\mathcal{L}_R + \delta_j^i (x^i - a^i)^{\alpha -1} \mathcal{L}_R.
\eea
Finally, if $\mathcal{T}_I^x$ is the infinitesimal rotation, we get the angular momentum tensor (AMT)
\bea\label{AMT}
\mathcal{M}^{\alpha, ik}_{j, R} =   -  x^k   \partial_j \phi({\bm x}) \, \partial_{\,_{}^a\D_i^{\alpha}\phi}\mathcal{L}_R +    x^k(x^i - a^i)^{\alpha -1}\delta^i_j\mathcal{L}_R.
\eea 
\end{proposition}

\begin{proposition}\label{prop2}
(Noether theorem) Assuming that the action \eqref{mamequann} is invariant under $\mathcal{T}_\mathcal{G}$ i.e. $\mathcal{T}_\mathcal{G} \mathcal{S}_L^\alpha-\mathcal{S}_L^\alpha=0$ for all $x^i\leq b^i$. Then there exists a globally conserved vector current given by
\bea\label{CourantL}
\Theta^{\alpha, i}_{(\sigma)L} &=&  C_{(\sigma)}({\bm x})\partial_{\D_i^{\alpha, b}\phi}\mathcal{L}_L +  \sum_{k =1}^{D}f^k_{(\sigma)}({\bm x})  \partial_k \phi({\bm x})\,  \partial_{\D_i^{\alpha, b}\phi}\mathcal{L}_L \cr
&&+  (x^i - b^i)^{\alpha -1} f^{i}_{(\sigma)}({\bm x})\,\mathcal{L}_L
\eea
In the case where the transformation $\mathcal{T}_I^x$ is the infinitesimal translation of vector ${\bm \epsilon}$ this current is reduced to the EMT
\bea\label{EMT}
\mathcal{T}^{\alpha, i}_{j,L} = -  \partial_j \phi({\bm x})\, \partial_{\,_{}^a\D_i^{\alpha}\phi}\mathcal{L}_L - \delta_j^i (b^i - x^i )^{\alpha -1} \mathcal{L}_L.
\eea
Finally, if $\mathcal{T}_I^x$ is the infinitesimal rotation, we get the AMT
\bea\label{AMT}
\mathcal{M}^{\alpha, ik}_{j, L} =   -  x^k   \partial_j \phi({\bm x}) \, \partial_{\D_i^{\alpha, b}\phi}\mathcal{L}_L -    x^k(b^i - x^i)^{\alpha -1}\delta^i_j\mathcal{L}_L. 
\eea 
\end{proposition}
\bigskip

The rest of this section is devoted to the proof of these propositions. The proof of proposition \eqref{prop1} will be given clearly, and the proof of proposition 2 follows logically. Denoted by $\mathcal{S}_R^{'\alpha}:=\mathcal{T}_I \mathcal{S}_R^\alpha$, we write
\begin{equation}
\mathcal{S}_R^{'\alpha}=\int \Delta_{R}^\alpha({\bm x'}) d^D{\bm x}'\mathcal{L}'(\phi'({\bm x}'),\,  _{}^{ a}\D^{'\alpha}\phi'({\bm x}')),
\end{equation}
and we then need to compute independently $\Delta_{R}^{\alpha}({\bm x}')$, 
 $d^D{\bm x}'$ and $\mathcal{T}_I\mathcal{L}:=\mathcal{L}'$. We start by
  expressing $(\Delta_{R}^{\alpha})^{-1}({\bm x}'):=\prod_{k=1}^D (x'^k - a'^k)^{1-\alpha}$
   using the transformation 
\eqref{tj} and by imposing the initial condition  $f_{(\sigma)}^k({\bm a})=0$, i.e. the vector ${\bm a}$ is invariant under the transformation $\mathcal{T}_I^x$ to ensure the integrability of the function $f_{(\sigma)}^k$ we get: 
\begin{align}
&(\Delta^\alpha_{R})^{-1}({\bm x'})\cr
&= \Big[\prod_{k=1}^D \Big( x^k - a^k  - \sum_{\sigma=1}^Mf^k_{(\sigma)}({\bm x})\beta^{(\sigma)}  \big)\Big]^{1-\alpha} \cr
&= (\Delta^\alpha_{R})^{-1}({\bm x})\Big[\prod_{k=1}^D  \Big(1- \sum_{\sigma=1}^M \beta^{(\sigma)}\frac{ f^k_{(\sigma)}(x)} {x^k - a^k }\Big)\Big]^{1-\alpha}\cr
&= (\Delta^\alpha_{R})^{-1}( x) \Big(\prod_{k=1}^D  \Big[1- (1-\alpha)\sum_{\sigma=1}^M\beta^{(\sigma)}\frac{ f^k_{(\sigma)}(x)} {x^k - a^k }\Big]\Big)\cr
&= (\Delta^\alpha_{R})^{-1}({\bm x})\Big[1 -(1-\alpha)\sum_{k=1}^{D}\sum_{\sigma=1}^M \beta^{(\sigma)}\frac{ f^k_{(\sigma)}( x)} {x^k - a^k } \Big].
\end{align}
Then by reversing and taking into account only the first order in $\beta$ we have
\beq\label{puting}
\Delta_R^\alpha({\bm x}') = \Delta_R^\alpha({\bm x})\Big[1 +(1-\alpha)\sum_{k=1}^{D}\sum_{\sigma=1}^M \beta^{(\sigma)}\frac{ f^k_{(\sigma)}({\bm x})}{x^k - a^k } \Big].
\eeq
It is simple to show that the measure $d^D{\bm x}$ is transformed as (we have used the identity $\det A=\exp[\tr \ln A]$ to re-express the Jacobian):
\bea\label{same} 
d^D{\bm x}' &=& \Big(1 - \sum_{i = 1}^{D}\sum_{\sigma=1}^M\partial_i f^{i}_{(\sigma)}({\bm x})\beta^{(\sigma)} \Big) d^D {\bm x}.
\eea
The transformation of the Lagrangian denoted by $\mathcal{L}_R'(\phi'({\bm x}'),_{}^a\D'^{\alpha}\phi'({\bm x}')),$ (where $_{}^a\D'^{\alpha}$ is the right fractional derivative at the point ${\bm x}'$) is derived by following the steps below. First let us recall that, the relation
\eqref{i1} we get:
\bea\label{pp}
_{}^a\D_i^{'\alpha}\phi(x'^{i}, {\bm x}'_{\bot_i}) &=& (x'^{i} - a'^i)^{1-\alpha} \partial_i^{'} \phi(x'^{i}, {\bm x}'_{\bot_i})\cr
&=& ( x'^i - a'^i)^{1-\alpha}\sum_{k =1}^{D}\Big(\frac{\partial x'^i}{\partial x^k} \Big)^{-1}\frac{\partial \phi'({\bm x'})}{\partial x^k},\cr
&&
\eea
with  $\frac{\partial x'^i}{\partial x^k} 
 =\delta^i_k - \sum_{\sigma=1}^M\partial_k f^i_{(\sigma)} \beta^{(\sigma)}  + O(\beta^2).
 $  Now using the expression of $x'^i$ and $\phi'(x')$ in the relation \eqref{t} we come to 
 \bea\label{}
_{}^a\D_{i}^{'\alpha}\phi'({\bm x}') = \,_{}^a\D^\alpha_i \phi({\bm x}) + \sum_{\sigma=1}^M  A_{i {(\sigma)}}^\alpha({\bm x})\beta^{(\sigma)} + O(\beta^2),
\eea
where the additional term coming from the transformation $\mathcal{T}_I$ is
\bea 
A_{i,{(\sigma)}}^\alpha({\bm x})& =&  \,_{}^a \D^\alpha_i C_{(\sigma)}({\bm x})  +  \sum_{\nu =1}^{D}\,_{}^a \D^\alpha_i f^\nu_{(j)}({\bm x})  \partial_\nu \phi({\bm x}) \cr
&&- (1-\alpha)\frac{f^i_{(\sigma)}({\bm x})  }{x^i - a^i}\, _{}^a\D^\alpha_i \phi({\bm x}).
\eea
\begin{widetext}
Then we can compute $\mathcal{L}_R'$ in the last step using the Taylor expansion, we get:
\bea\label{Lagg}
\mathcal{L}_R'&=& \mathcal{L}_R\Big( \phi({\bm x}) +  \sum_{\sigma=1}^M C_{(\sigma)}({\bm x})\beta^{(\sigma)}, \, \,_{}^a\D^\alpha \phi({\bm x}) +  \sum_{\sigma=1}^M A_{\cdot{(\sigma)}}^\alpha({\bm x})\beta^{(\sigma)}\Big) \cr
&=& \mathcal{L}_R + \sum_{\sigma=1}^M C_{(\sigma)}({\bm x})\beta^{(\sigma)} \partial_\phi \mathcal{L}_R  + \sum_{i =1}^{D}\sum_{\sigma=1}^M  A_{i {(\sigma)}}^\alpha({\bm x})\beta^{(\sigma)}\partial_{\,_{}^a\D_i^{\alpha}\phi} \mathcal{L}_R   + O(\beta^2).
\eea
The transform action $\mathcal{S}'^\alpha_L$ may be deduced using \eqref{puting}, \eqref{same} and \eqref{Lagg} as
\bea\label{varg}
\mathcal{S}'^\alpha_R& =& \int_{\mathcal{D}_R^\alpha} \, \Delta^\alpha_{R}({\bm x}) d^D {\bm x}  \Bigg[\mathcal{L}_R  + \sum_{\sigma=1}^M C_{(\sigma)}({\bm x})\beta^{(\sigma)} \partial_\phi \mathcal{L}_R +  \sum_{i =1}^{D} \sum_{\sigma=1}^M \Bigg ( A_{i {(\sigma)}}^\alpha({\bm x})\beta^{(\sigma)}\partial_{\,_{}^a\D_i^{\alpha}\phi}\mathcal{L}_R \cr 
&+& (1-\alpha)\beta^{(\sigma)}\frac{ f^i_{(\sigma)} ({\bm x})  }{x^i - a^i} \mathcal{L}_R - \partial_i f^{i}_{(\sigma)}({\bm x})\beta^{(\sigma)}\mathcal{L}_R\Bigg)\Bigg].
\eea
Finally, the variation of the action under the infinitesimal transformation $\mathcal{T}_I$ replacing the expression of  $ A_{i {(\sigma)}}^\alpha$ given in \eqref{varg} and after few algebraic computations is
\bea\label{18g}
_{}^a\delta^\alpha\mathcal{S}_R^\alpha& =&\sum_{\sigma=1}^M \beta^{(\sigma)} \int_{\mathcal{D}_R^\alpha} \Delta^\alpha_{R}({\bm x})d^D {\bm x}   \Bigg\{\sum_{i =1}^{D}\,_{}^a\D_i^\alpha\, \Big(\Theta^{\alpha,i}_{(\sigma)R}\Big) + \mathcal{B}_{ (\sigma)R}\Bigg\},
\eea
where $\Theta^{\alpha, i}_{(\sigma)R}$ and $\mathcal{B}_{ (\sigma)R}$ are given respectively by
\bea\label{moi}
\Theta^{\alpha, i}_{(\sigma)R} &=&  C_{(\sigma)}({\bm x})\partial_{\,_{}^a\D_i^{\alpha}\phi}\mathcal{L}_R +  \sum_{k =1}^{D}f^k_{(\sigma)}({\bm x})  \partial_k \phi({\bm x})\,  \partial_{\,_{}^a\D_i^{\alpha}\phi}\mathcal{L}_R -  (x^i - a^i)^{\alpha -1} f^{i}_{(\sigma)}({\bm x})\mathcal{L}_R\\
\mathcal{B}_{ (\sigma)R} &= & \sum_{i=1}^{D}\Bigg[-\sum_{k=1}^D  f^k_{(\sigma)}({\bm x}) \,_{}^a\D^\alpha_i\Big( \partial_k \phi{(\bm x)} \, \partial_{_{}^a\D_i^\alpha\phi}\mathcal{L}_R \Big) - (1-\alpha)\frac{f^i_{(\sigma)}({\bm x})  }{x^i - a^i} \D^\alpha_i \phi   \partial_{_{}^a\D_i^{\alpha}\phi}\mathcal{L}_R +\cr
& +&   (1-\alpha)\frac{ f^i_{(\sigma)}({\bm x})}{x^i - a^i} \mathcal{L}_R + f^{i}_{(\sigma)}({\bm x})\,_{}^a\D_i^\alpha\Big((x^i - a^i)^{\alpha -1}\mathcal{L}_R\, \Big) + C_{(\sigma)}\Big( \partial_\phi \mathcal{L}_R - \,_{}^a\D^\alpha_i\big( \partial_{_{}^a\D^\alpha_i \phi} \mathcal{L}_R \big) \Big) \Bigg] \label{malboro}.
\eea
\end{widetext}
A few computations allow us to conclude that the broken term provided by our analysis is identically zero. To convince oneself of this assertion let us separate the terms in  \eqref{malboro} as 
k
Due to the EL equation of motion, this expression is reduced to
\begin{align}
\mathcal{B}_{ (\sigma)R}
&=  - (1-\alpha)\sum_{i =1}^{D} \frac{f^i_{(\sigma)} ({\bm x})  }{x^i - a^i} \,_{}^a\D^\alpha_i \phi (\bm{x})  \partial_{\,_{}^a\D_i^{\alpha}\phi}\mathcal{L}_R \cr
&- \sum_{i, k =1}^{ D}f^k_{(\sigma)}({\bm x})  _{}^a\D^\alpha_i\big( \partial_k \phi  \big) \partial_{_{}^a\D_i^{\alpha}\phi}\mathcal{L}_R + \cr 
& + \sum_{i, k =1}^{ D}f^k_{(\sigma)}({\bm x}) \partial_k (\,_{}^a\D_i^\alpha \phi)\partial_{\,_{}^a\D_i^{\alpha}\phi} \mathcal{L}_R  .
\end{align}

Now using the following identity
$
\partial_k (_{}^a\D_i^\alpha \phi) =\frac{(1-\alpha)}{(x^k - a^k)^\alpha}\delta^k_i \partial_i\phi + \,_{}^a\D^\alpha_i(\partial_k\phi)
$
we finally get $\mathcal{B}_{(\sigma)R} =0$.\\

As an example of Noether current, let us assume that the transformation $\mathcal{T}_I$ is the infinitesimal spacetime translation of vector ${\bm \epsilon}=(\epsilon^1,\epsilon^2,\cdots, \epsilon^D)$. This implies that the group parameter dimension is exactly the spacetime dimension $D$ i.e. $M=D$ and we write
\bea\label{pok}
\begin{cases}
  x'^i &=x^i + \epsilon^i \cr 
 \phi'({\bm x}') &= \phi({\bm x})
\end{cases}
\quad\Rightarrow
\begin{cases}
\delta  x^i& =  \epsilon^i \cr 
\delta \phi({\bm x})& = 0.
\end{cases}
\eea
Note that the relation
\eqref{pok} is the same with \eqref{tj} by identified  $\sum_{\sigma=1}^Nf_{(\sigma)}^i\beta^{(\sigma)} \equiv -\, \epsilon^i $ and $\sum_{\sigma=1}^N \beta^{(\sigma)}C_{(\sigma)}({\bm x}) \equiv 0$. The EMT \eqref{EMT} can be simply obtained using this prescription. In the same manner, assuming that the transformation $\mathcal{T}_I$ is the spacetime rotation:
\beq\label{125}
\begin{cases}
	 x'^i &= x^i - \sum_{k=1}^{D}\omega^i_k x^k  \cr 
	 \phi'({\bm x}')&= \phi( {\bm x})
\end{cases}\Rightarrow
\begin{cases}
	\delta x^i &= - \sum_{k=1}^{D}\omega^i_k x^k\cr
	\delta \phi &= 0.
\end{cases}
\eeq
where $[\omega {\bm x}]:=(x^1-\sum_{k=1}^D\omega_k^1x^k,\cdots,x^D-\sum_{k=1}^D\omega_k^D x^k)$. Note that we identified 
the syst\`eme \eqref{125} to the transformation \eqref{tj} as $\sum_{\sigma=1}^Nf_{(\sigma)}^i\beta^{(\sigma)} \equiv - \sum_{k=1}^{D}\omega^i_k x^k$ and $\sum_{\sigma=1}^N C_{(\sigma)}(\bm{x})\beta^{(\sigma)} \equiv 0$ and therefore the AMT \eqref{AMT} may be simply obtained.

Let us now come to the following remark. The Noether current \eqref{CourantR} and \eqref{CourantL} representing the right and left parts will be added in the ${\bm a}\rightarrow {\bm b}$ limit to cover the general Noether current. Then we have
\bea
\Theta^{\alpha, i}_{(\sigma)}=\Theta^{\alpha, i}_{(\sigma)R}+\Theta^{\alpha, i}_{(\sigma)L} \Big|_{{\bm a}\rightarrow {\bm b}}
\eea
In the same manner, we get
$
\mathcal{T}^{\alpha, i}_{j}=\mathcal{T}^{\alpha, i}_{j,R}+\mathcal{T}^{\alpha, i}_{j,L}\Big|_{{\bm a}\rightarrow {\bm b}}$ and 
$
\mathcal{M}^{\alpha, ik}_{j}=\mathcal{M}^{\alpha, ik}_{j, R}+\mathcal{M}^{\alpha, ik}_{j, L}\Big|_{{\bm a}\rightarrow {\bm b}}
$.
\section{Application to one dimension harmonic oscillator }\label{sec4}
\begin{figure}
\begin{center}
\includegraphics[scale=0.4]{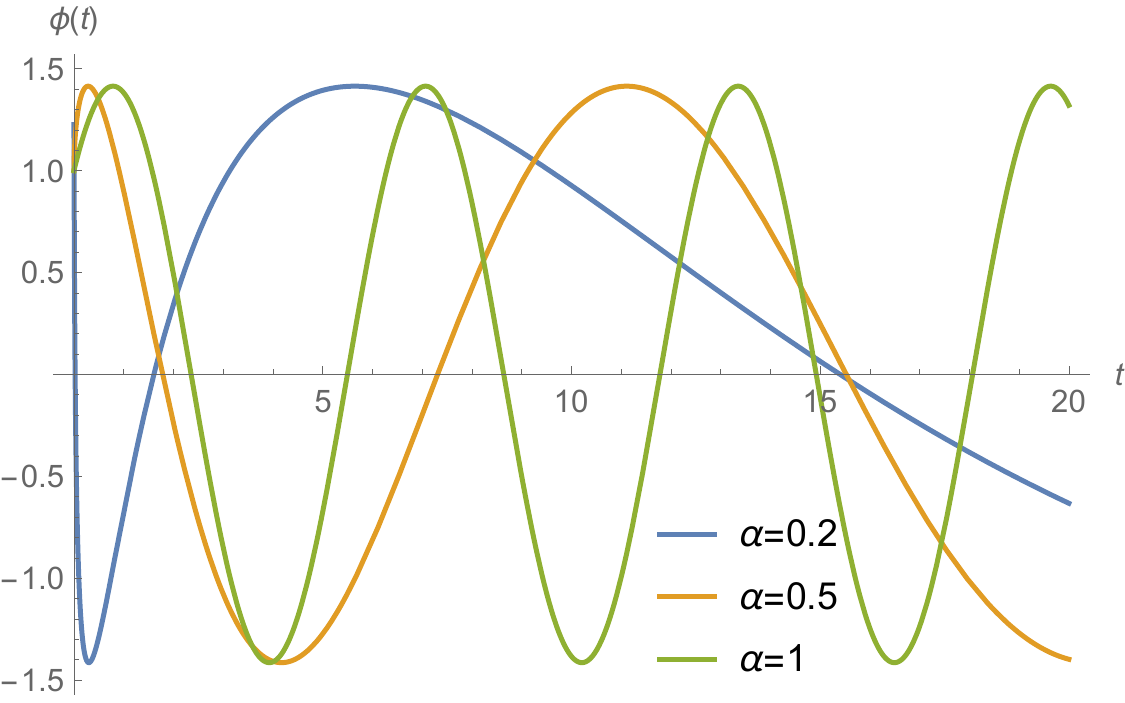}
\end{center}
\caption{Representation of the harmonic oscillator solution for three particular values of $\alpha$. For $\alpha=1$ we find the result of the classical harmonic oscillator. But for other values of  $\alpha$ away from $1$,  we see that this is a delayed oscillator. We have used the initial condition $\phi[(\frac{2\pi\alpha}{m})^{\frac{1}{\alpha}}]=1$ and $\frac{d}{dt}\phi[(\frac{2\pi\alpha}{m})^{\frac{1}{\alpha}}]=1$, $m=1$.}\label{jpgraphe}\label{fig2}
\end{figure}

In this section, we apply our analysis to a simple problem such as a harmonic oscillator in this fractional spacetime. Particularly, we reduce the dimension to one for simplicity, which is also represented by the temporal coordinate denoted by $t:=x^1$. To respect the dimensional analysis in the definition of the Lagrangian, we will impose a parameter $\xi_d$  multiplying the kinetic term. Thus, we have 
\bea
\mathcal{L}_R &=& \frac{1}{2}\xi_d(\,_{}^a\D^\alpha_t\phi(t))^2 -\frac{1}{2}m^2 \phi^{2}(t), \quad t\geq a\\
\mathcal{L}_L &=& \frac{1}{2}\xi_d(\D^{\alpha, a}_t\phi(t))^2 -\frac{1}{2}m^2 \phi^{2}(t),\quad t\leq a.
\eea
Once again, we set $a=0$. The EL equation of motion \eqref{xxxx} and \eqref{xxxx1} are expressed as 
\beq\label{HO1} 
\Big[(1-\alpha)t^{1-2\alpha}\partial_t  + t^{2-2\alpha}\partial^2_t  + \widetilde{m}^2\Big]\phi  = 0,\,\, t\geq 0
\eeq
\beq\label{HO2}
\Big[(1-\alpha)(- t)^{1-2\alpha}\partial_t  + ( - t)^{2-2\alpha}\partial^2_t   + \widetilde{m}^2\,\Big]\phi  = 0,\,t\leq 0,
\eeq
where $\widetilde{m}^2=\frac{m^2}{\xi_d}$
The limit $\alpha=1$ corresponds to $\xi_d=1$ and we recover the harmonic oscillator equation of motion. Let us now remark also that equation \eqref{HO2} is symmetric to \eqref{HO1} by time reversal $t\rightarrow -t$.  Therefore, we can solve \eqref{HO1} as: (Let $A,B\in\mathbb{R}$ are two integration constants), then
\beq\label{physicalsol}
\phi(t)=A\cos\frac{\widetilde{m}t^\alpha}{\alpha}+B\sin\frac{\widetilde{m}t^\alpha}{\alpha},\,\,\forall t\geq 0.
\eeq
By replacing $t$ by $-t$ to recover the negative solution, we conclude that $\phi(t\leq 0)$ is not real and can not be considered in our analysis here. The physical solution is only given by \eqref{physicalsol}. It should therefore be noted that solving the EL equation of motion fixes the domain of validity of the Noether current and in the harmonic oscillator case is reduced to $t\in[0,\infty[$.
In figure \eqref{fig2} we represent the solution \eqref{physicalsol} and note the loss of local conservation of the EMT for values of $\alpha$ away from $1$.

The computation of the oscillator energy in this case corresponds to  $\mathcal{E}_R = -\mathcal{T}^{\alpha, 1}_{1, R}$ given by
\bea
\mathcal{E}_R  &=&   \partial_t\phi \,\partial_{_{}^a\D_t^\alpha \phi }\mathcal{L}_R - t^{\alpha -1}  \mathcal{L}_R \cr
&=&  \frac{1}{2}\xi_d t^{1-\alpha}(\partial_t \phi)^2 + \frac{1}{2}t^{\alpha -1} m^2\phi^2.
\eea
Then the partial derivative of $\mathcal{E}_R$ with respect to $t$ by considered the solution \eqref{physicalsol} is 
\beq
\frac{d\mathcal{E}_R}{dt}=\frac{1}{2} (\alpha -1) m^2  (A^2+B^2) t^{\alpha -2},
\eeq
which shows that the conformable fractional oscillator is not a conserved system for $\alpha\neq 1$ and is following the diagram \eqref{fig2}. Finally, using the usual regularization method of the EMT $\mathcal{E}_R$ can be regularized to $\widetilde{\mathcal{E}}_R$ given by
\beq
\widetilde{\mathcal{E}}_R=\mathcal{E}_R-\frac{1}{2} m^2  \left(A^2+B^2\right) t^{\alpha -1},
\eeq
which is now locally conserved. The same procedure can be applied to the general expression of the EMT to study its regularization. This analysis is deserved for our future investigation of this subject.

\section{Conclusion and outlooks}\label{sec5}
In this work, we constructed a Lagrangian approach for the dynamics of a scalar field on a conformable fractional spacetime, i.e. where the standard notion of derivation is replaced with a fractional derivative. This first step was concluded by investigating the fractional equations of motions for a free scalar field. Note that, despite we focused on a scalar field, the construction we proposed could be easily extended for other kinds of fields. 

Before defining the Lagrangian approach from a variational point of view, we move on to the study of the relation between symmetries and conservation laws in this setting, generalizing the classical Noether's theorem.  We have considered global symmetries of the action supported by some group of symmetry acting both on internal and external variables. A crucial point for the derivation was the cancellation of the breaking term $\mathcal{B}_{\sigma}$, which appears as a pure consequence of the fractional derivative (i.e. does not appear for  $\alpha\to 1$) and which vanishes ‘‘on shell", as soon as the field satisfy the equations of motion. Finally, we derived the generalization of EMT and AMT, which are generally considered in field theories to address conservation laws.

We expect that the formulation of field theories on deformed spaces and further on spaces with fractional derivatives are promising frameworks for the exploration of high energy physics beyond the standard model, where small distance space-time is expected to lack some standard properties of low energy regime. Our analysis indicates that the conformable derivative is a good definition of the fractional derivative to construct field theory based on a variational principle. In modern physics furthermore, variational principles are expected to come, in fact, as the classical limit of path integrals defining the quantized field theories. Hence, we expect such a path-integral approach can be defined in this framework, and we plan to investigate path-integral quantization in a forthcoming work. This, indeed, is only the first step of a program aiming to investigate the renormalization group, symmetry breaking and Higgs mechanism for a deformed version of the standard model, with the final goal to provide some phenomenological predictions eventually indicating that such a fractional derivative may be an effective tool to address high energy phenomena. 

 Finally, let us comment on the hypothesis that allowed the analytical computation of this work.  We focused on a restricted family of trajectories, for which ordinary derivatives exist. This restriction loses the enriched structure coming from fractional space leading to the trajectories which are not derivable in the ordinary sense but allow using of the correspondence \eqref{corresp}. Giving up this hypothesis increases the computation difficulty, and we plan to address this challenging issue in a forthcoming work.

\section*{Acknowledgements} 
V.L. would like to thank the little crab for his inspiration in the final stages of this work.



\onecolumngrid
  
\end{document}